# A simple model of cognitive processing in repeated games

Anders Eriksson and Kristian Lindgren
Physical Resource Theory – Complex Systems, Department of Energy and Environment,
Chalmers University of Technology, SE-41296, Göteborg, Sweden
anders.eriksson@chalmers.se, kristian.lindgren@fy.chalmers.se

**Abstract**

In repeated interactions between individuals, we do not expect that exactly the same situation will occur from one time to another. Contrary to what is common in models of repeated games in the literature, most real situations may differ a lot and they are seldom completely symmetric. The purpose of this paper is to discuss a simple model of cognitive processing in the context of a repeated interaction with varying payoffs. The interaction between players is modelled by a repeated game with random observable payoffs. Cooperation is not simply associated with a certain action but needs to be understood as a phenomenon of the behaviour in the repeated game. The players are thus faced with a more complex situation, compared to the Prisoner's Dilemma that has been widely used for investigating the conditions for cooperation in evolving populations. Still, there are robust cooperating strategies that usually evolve in a population of players. In the cooperative mode, these strategies select an action that allows for maximizing the sum of the payoff of the two players in each round, regardless of the own payoff. Two such players maximise the expected total long-term payoff. If the opponent deviates from this scheme, the strategy invokes a punishment action, which aims at lowering the opponent's score for the rest of the (possibly infinitely) repeated game. The introduction of mistakes to the game actually pushes evolution towards more cooperative strategies even though the game becomes more difficult.

**Introduction**

In repeated interactions between individuals, we do not expect that exactly the same situation will occur from one time to another. Contrary to what is common in models of repeated games in the literature, most real situations may differ a lot and they are seldom completely symmetric. How much cognitive sophistication is needed to deal with such situations? The purpose of this paper is to discuss a simple model of cognitive processes in the context of a repeated interaction with varying payoffs.

One of the phenomena we will discuss in an evolutionary perspective is the emergence of cooperation in a population. In the repeated game where payoffs are varying, cooperation cannot be directly associated with a certain action, as it can be in the Prisoner's Dilemma (PD) game (Flood 1958). In that way, the PD game actually avoids the problem of defining, in a broader sense, what cooperation means in a repeated interaction. Cooperation in biological and social systems can often appear as if the player place the good of the group (or species) before immediate personal gain, contrary to what one would expect for behaviour evolved under natural selection or individual profit maximisation. It is well known, though, and it has been demonstrated in the literature, e.g., originating from the work of Trivers (1971), that seemingly altruistic or cooperative behaviour can be in the self-interest of the individual (or her genes) in a variety of different situations.

A large body of theoretical (and numerical) analysis, focusing on cooperation in an evolutionary context, has been developed during the past two decades using the PD as the paradigm for interaction. The PD is attractive as a simple model of the dilemma faced in many situations when shortsighted profit-maximisation is a tempting choice instead of a shared reward, yielding higher long-term profits if cooperation can be established. An example in a social system is given by the "tragedy of the common", where individuals that harvest more from a common resource may jeopardize cooperation expressed as an agreed sustainable level of resource extraction (Hardin 1968; Ostrom 1980).

The possibility for a cooperative equilibrium to be established in a repeated game has long since been well known in game theory. The folk theorem states that in a repeated game, with sufficiently low probability for the game to end, any possible score above the min-max payoff can be supported at equilibrium by some strategy (Fudenberg and Tirole 1991; Binmore 1994). Dutta (1995) extended this analysis to stochastic, and possibly very complex game dynamics, when the number of actions and payoff values are bounded. Such equilibria are kept by punishing those who deviate from the equilibrium strategy. One example of punishment is to minimise the possible score for the opponent in the present round. However, the folk theorem does not say anything on if, how, or which equilibrium is reached. Such questions are the targets for evolutionary models of the game-theoretic problem in question.

The work by Axelrod (Axelrod and Hamilton 1981; Axelrod 1984, 1987) was a starting point for a series of papers presenting an evolutionary perspective on how cooperation can be established. The main conclusion from the initial work was that cooperation could be established if, first, interactions between individuals are repeated, and, second, the number of interactions has a stochastic component. Usually, it is assumed that the game is discontinued with a probability $r$ after each round, leading to a geometrically distributed number of rounds. A large number of modifications and extensions of the PD was tried to firmly establish the fact that cooperation is possible under a wide variety of circumstances (Matsuo 1985; Molander 1985; Axelrod 1987; Boyd and Lorberbaum 1987; Boyd 1989; Miller 1989; Lindgren 1992; Nowak and May 1993a,b; Stanley *et al*. 1993; Ikegami 1994; Lindgren and Nordahl 1994; Nowak *et al*. 1995; Wu and Axelrod 1995; Lindgren 1997).

The general assumption in game theory is that individuals behave rationally in the sense that they always choose the action (if possible) that maximise the individual score. This assumption relies on the idea that games can be viewed as isolated events and that an individual's behaviour can be considered separately from the broader social context. In experimental work it has been repeatedly demonstrated that human behaviour is more cooperative than what is dictated from the rationality assumption in game theory (Heinrich *et al*. 2001; Fehr and Gächter 2000a,b, 2001). This fact is important to bear in mind when making simple models and interpreting their results.

### *Characterising two-player games*

The study of game theory has given rise to a wealth of models for describing different situations, where the PD is but one example. In a symmetric two-player game, with two actions (labelled 1 and 2), it is always possible to bring the payoff matrix $u$ to a standard form where the diagonal entries are $u_{11} = 1$ and $u_{22} = 0$, and the off-diagonal entries are $u_{12} = S$ and $u_{21} = T$, with a transformation that preserves the properties of the game under, e.g., the replicator dynamics. In the literature, the values of the off-diagonal elements $T$ and $S$ has been partitioned into regions with different properties. As is illustrated in Figure 1, there are several situations, in which two individuals may interact, with characteristics completely different from the PD game. Each region is often illustrated by a colourful story, pertaining to a certain type of social situation.

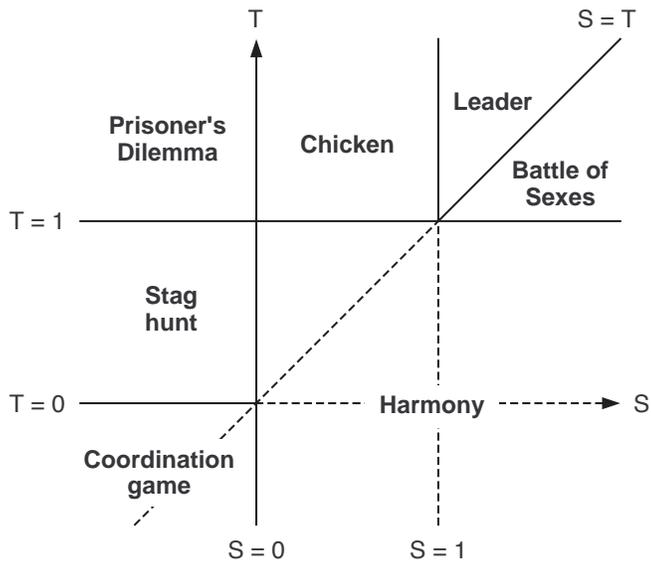

**Figure 1:** In a symmetric two-player game, with two actions, it is always possible to reduce the payoff matrix to a matrix where the elements on the diagonal is 1 and 0, with a transformation that preserves the properties of the game under, e.g., the replicator dynamics. The figure shows how different values of the parameters *S* and *T* correspond to models in the game theory literature.

In the PD game, players can either cooperate or defect. The reward for cooperation is 1. $T > 1$ is the temptation score (to defect against a cooperator), 0 is the punishment score when both defect, and $S < 0$ is the payoff to a cooperator when the opponent defects. (One usually also have $S + T < 2R$, so that cooperation is favoured over taking turns defecting in the iterated game.) This game has a unique pure strategy Nash equilibrium with the mutual score 0. In the Battle of Sexes (corresponding to $1 < T < S$) there are two pure strategy Nash equilibria[1] given by the payoff pairs $(T, S)$ and $(S, T)$. Both of the games Chicken (corresponding to $0 < S < 1 < T$) and Leader (corresponding to $1 < S < T$) (Guyer and Rapoport 1966) have the same pure strategy equilibria, but they capture different types of situations. If put in an evolutionary context and allowing for repeated interactions one may observe very different behaviour evolving in these games. In Stag hunt (corresponding to $S < 0 < T < 1$) both payoff pairs (1,1) and (0,0) are Nash equilibria, but obviously one is the best for both. It is worth to notice that many situations may actually be fairly well described with both T and S being in the interval [0, 1], often referred to as the harmony game, and in that case (1,1) is both a Nash equilibrium and the best outcome for both.

The picture becomes even more complicated if the constraint on symmetry is released. Asymmetric games, especially games where the payoffs are symmetrical on average, may be used to model situations where costly signalling is important (Hauser 1996).

One of the main limitations with the PD game and many of the similar games studied is the static character of the interaction situation. Each time two individuals encounter each other in the repeated PD game the situation is identical to the previous one, i.e., the payoff values for the different choices are unchanged. In a real situation, it may be much more common that they meet each other in different situations most of the time; some situations may be of the PD type, other situations may be more straightforward, so short-sighted profit maximisation coincides with the cooperative maximal payoff, and even more importantly, many situations may be asymmetric. Even when the same people meet each other, they rarely meet under precisely the same circumstances.

---

[1] In this paper, we only consider pure strategies, i.e., there is a deterministic choice of action in every given situation. A Nash Equilibrium is then a pair of actions of the players, such that none of them would be better off switching to the other action if the opponent stays with her choice.

Johnson *et al* (2002) argue in a similar way. Starting with a PD payoff matrix they illustrate how payoff variations, if sufficiently large, frequently violates the conditions on the payoff matrix that defines the PD game. They use this observation as an explanation for why PD situations are so rarely found in natural systems.

The advantage with the PD game model is its simplicity, allowing for various extensions suitable for investigating different characteristics of cooperative behaviour in simple models. Our aim with the current research is to keep the simplicity at the same time as the situations in which agents meet are less static. Therefore, we want to argue for a more general model of the situation in pairwise interactions in which payoff values are changing from one encounter to the next and where there is no constraint on symmetry. In our model, we generate a completely new random payoff matrix for each round in a repeated two-player game. Each player chooses between two actions and has full information about the payoff matrix. The matrix consists of four randomly (and independently) generated payoff pairs corresponding to the pair of action choices by the players. This means that in each round, the players face a completely new situation.

Some effect of random payoff values can be illustrated already in the context of symmetric games in the S–T plane of Figure 1. There are 12 different regions (two of which are occupied by the Coordination game), corresponding to the different orderings of the four scores in the payoff matrix of the game. If the score values are independent and identically distributed, it follows that, for any distribution, the total probability of each region equals 1/12. Thus, it is equally probable to have a matrix corresponding to a PD as to any of the other games (note, however that Harmony occurs with probability 5/12 because it covers five regions). In our model, the symmetry constraint is removed, and we assume that all payoff values are independently and uniformly distributed in the unit interval [0, 1]. Figure 2 shows the S–T distribution of the payoff matrix of one of the players, and illustrates the spread in characteristics of the game (c.f. Figure 1) that occurs due to the stochastic generation of payoff elements.

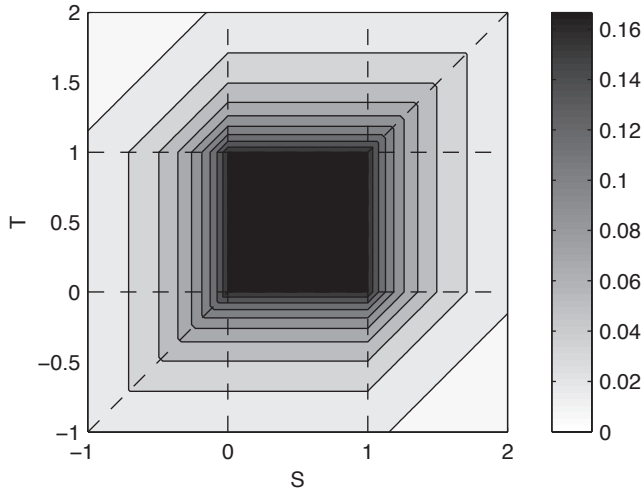

**Figure 2:** In our model, the payoff matrix elements are independent and uniformly distributed random numbers. The corresponding distribution function of the parameters $S$ and $T$, in the standard form of each matrix, can be calculated analytically to be $(\max\{S,T,1\} - \min\{S,T,0\})^{-3}/6$ (Eriksson, 2004, unpublished). Compare to Figure 1, which gives an overview of how games in the literature correspond to regions in the S–T plane. As can be seen in the figure, it is piecewise polynomial in each of the twelve regions, and is continuous in the whole plane.

One of the main points of this paper is to present and argue for a simple model that captures a sufficiently wide range of behaviours relevant for this type of repeated interaction. Since there is no specific meaning associated with the two actions that the player can choose from in this game, like the interpretation of cooperation and defection in the PD game, we need the players to be capable of interpreting the action of their opponent in the context of the current payoff matrix. Therefore, we have chosen a limited set of basic behaviours that a player can choose from, which determines the action depending on the current payoff matrix. These behaviours include seeking a Nash equilibrium (a pair of actions from which no player has incentive to deviate unilaterally), minimising the opponent's maximum score, and aiming for maximum of the sum of both player's scores. After observing the action of the opponent, a player can interpret the behaviour of the opponent in terms of these basic behaviours. If the player now has a higher-level strategy, the result of this interpretation may be a change in the player's own basic behaviour for the next round. Some preliminary results from tests of this idea have been reported elsewhere (Eriksson and Lindgren 2001, 2002). The composite (higher level) strategies described here, can be viewed as a very simple model for a cognitive process in a repeated game situation, which involves both the interpretation of the opponent's behaviour and a response to this by choice of basic behaviour in the next interaction.

One questions that immediately arises in this situation is: What does cooperation mean when we do not have the PD game structure - not even a specific action that can be called "cooperation"? Our main candidate for a definition of cooperation in a repeated game, as the one described above, is a sequence of pair of actions that results in the highest (equal) long-term score for both players. In this statement, there is an assumption that the games in the long term are symmetric and that 'equal' holds for the expectation values of the scores.

## Methods

The methods section is divided into three parts. First, we describe the basic behaviours that are available for the players. A basic behaviour dictates what action the player should take based on the observation of the current payoff matrix, i.e., it is a mapping from the current payoff matrix to the actions. We will restrict the model to a set of nine behaviours, which spans a large fraction of the space of joint expected payoffs available to pairs of players. This behavioural repertoire can be interpreted as the different "moods" of the player. Some behaviours are cooperative, others are competitive or selfish. Second, we describe how players build complex strategies from behaviours by changing behaviour conditional on the actions of the opponent. An important complication here is that for given payoff matrices, several behaviours may dictate the same action. The important cognitive component here is that a player interprets the action of the opponent in terms of the basic behaviours. Third, we describe the simple evolutionary model of how strategies evolve and spread in the population.

### *Basic behaviours*

The basic behaviours depend only on the present payoff matrices, and there is no memory of earlier interactions. In this case, one may analyse the situation as a single round game, averaging over all possible payoff matrices. Therefore, one reasonable first approach, as a basis for constructing a basic behaviour, is to look for Nash Equilibria (NE) in the current payoff matrix. If there is one NE, selfish profit-maximisation implies that one should choose the action that corresponds the NE. With our assumption of uniformly distributed payoff values between 0 and 1, it turns out that 75% of the matrices contain exactly one NE, while the remaining 25% of the cases are divided equally between zero and two NEs. As a first basic behaviour, we choose the strategy *NashSeek* that aims for the NE with the highest sum of both players score or, if no NE exists, optimistically aims for the highest possible own payoff. Another basic behaviour that is important in the repertoire is *Punish*, which is a strategy that minimises the maximum payoff of the opponent.

The third basic behaviour we consider is one that will play a crucial role in the evolution of cooperation in the repeated game. If two players have the same strategy, it is clear that the best they can do is to maximize the sum of both players' payoffs, regardless of their own score. On average, they will share equally the highest possible score, even though in some situations this means choosing a smaller payoff for the benefit of the opponent. We denote by *MaxCoop*, the strategy that aims for the maximum sum of both players' payoffs.

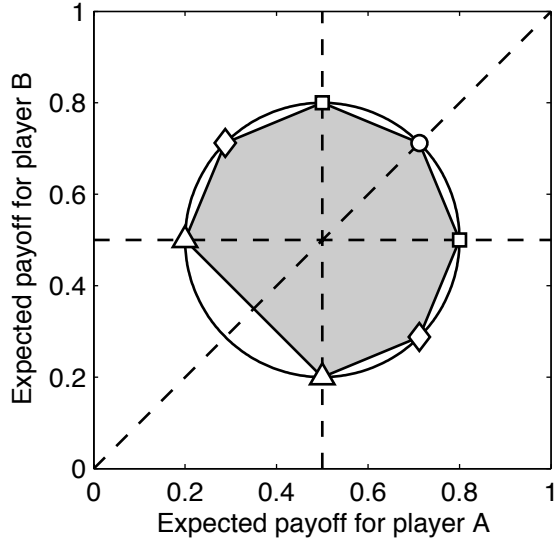

**Figure 3:** The range of joint expected payoffs is very close to a circle with radius 0.3, centred at (0.5, 0.5). The grey area corresponds to the set of attainable pairs of expected payoffs for strategies based on the nine basic behaviours. The symbols along the border correspond to pairs of basic behaviours: the circle corresponds to MaxCoop/MaxCoop, the squares to the Generous/Optimistic, the diamonds to Discriminate/AntiDiscriminate and the triangles to Sorry/Hurting. The expected payoff for MaxCoop against MaxCoop is 3589/5040 ≈ 0.71210, which illustrates that the true border is slightly flatter than a circle (the corresponding value for the circle is 0.71213).

The set of possible average scores that can be obtained by any pair of memory-less strategies[2] is shown in Figure 3. The edge, which is close to a perfect circle with radius 0.3 centered at (0.5, 0.5), is represented by pairs of different coordinated strategies. The furthest top-right point is the scores of two MaxCoop strategies. The score of two NashSeek players is located within the circle, at a smaller value than is MaxCoop. In order to have a set of basic behaviours that gives a reasonable cover of the possible average scores, we also include the following coordinated pairs of basic behaviours:

- Behaviours *Sorry* and *Hurting* jointly achieve the minimum possible expected payoff of 0.2 for the player that follows the Sorry strategy.
- *Generous* and *Optimistic* coordinate to give the Optimistic player the maximum possible expected payoff, 0.8.
- *Discriminate* and *AntiDiscriminate* maximise the difference in payoff, between the player and the opponent (Discriminate) and the converse (AntiDiscriminate).

---

[2] We generalise the principle behind the *MaxCoop* strategy to find the set of joint expected payoffs the players can have. Each point on the border corresponds to a tradeoff between the expected payoffs of the players. Formally, consider a line, originating at the pair of expected payoffs (0.5, 0.5) for two players picking either action with equal probability. The stochastic payoff matrices of players A and B are $u$ and $v$, respectively, and their respective actions are denoted $i$ and $j$. If the line extends at an angle $\theta$ to the x-axis, it intersects the boundary of the set of joint expected payoffs at the position $(\langle u_{i^*j^*}\rangle, \langle v_{i^*j^*}\rangle)$, where $(i^*, j^*) = \arg\max_{ij}(\cos\theta\, u_{ij} + \sin\theta\, v_{ij})$ and the angular brackets denotes expected value. Note that the players' strategies are deterministic; the payoffs are random variables only because the payoff matrices are stochastic.

### *Strategies for the repeated game*

In the repeated game, the game ends with probability $r$ after each round, so that the number of rounds is geometrically distributed with expected number of rounds $n = 1/r$. Thus, the parameter $r$ is the discount rate of the players.

The total space of possible strategies for the iterated game is very large, since one could in principle take into account the full history of payoff matrices and actions, as well as the current payoff matrix. In order to have a rich yet reasonably simple model, we divide the strategy space into a behavioural, memory-free part reacting on the payoff matrices in each round, and a strategic part that commits to one of the nine behaviours ahead of each round, and may switch to another behaviour in the next round depending on the outcome of the present round.

The finite-state automaton (FSA) is a very powerful representation of a strategy, capable of representing dependence on the history of the game – even far back – and complex transitions, while remaining simple and intuitive in its form. Several authors have used FSA representations in strategies for the PD (Akima and Soutchanski 1994; Miller 1996; Nowak *et al.* 1995; Lindgren 1997). An FSA consists of a certain number of internal states, each associated with an action and connected to some of the other states (called neighbours). In addition to the action associated with a state, the automaton has conditions that govern the transition from the present state to a neighbouring state, based on the outcome of the round. Thus, the history of actions up to the present round of the game determines the state of the automaton, but given this, the state in the next round depends only on actions in the present round. Along with the state of the FSA in the first round, actions and transition rules completely determines the behaviour of the FSA. We have extended this approach to the case of randomly varying payoff matrices. Instead depending directly on the action of the opponent, the transition rules now dictate that the player should compare the action of the opponent to the actions that the nine basic behaviours would give, and based on these comparisons move to a different state. For instance, a rule might require that the opponent's action is the same as that of MaxCoop, is opposite of NashSeek's, and ignore the other behaviours.[3]

In the remaining part of this section, we discuss our use of FSAs to calculate the expected score of a player. An attractive feature of the FSA representation of strategies is that a game between two players corresponds to a Markov process. A state of the Markov process corresponds to the pair of states of the players' strategies in a given round. This means that we can calculate all quantities of interest from the transition matrix of the process. In calculating the expected payoffs of two complex strategies, we proceed in two steps. First, we construct the transition probability matrix of the game from the FSAs of the players' strategies, which gives the probability of jumping from one Markov state to another in a round. In this calculation, we use a pre-calculated table of the relative frequency of different action profiles[4] (simultaneous actions of the nine basic behaviours for both players). The distribution over action profiles is calculated from the distribution of payoff elements and the nine behaviours. Second, the contribution from a given round $t$ (where $t = 1,2,...,\infty$) to the expected score of a the player is the product of the probability that the game lasts at least t rounds (given by $(1-r)^{t-1}$), and the expected payoff to the player in each Markov state (also pre-calculated), weighted by the probability distribution over the set of joint states at time $t$ (which can be calculated from the probability transition matrix and the initial state of both players). In order to be able to compare expected scores for different discount rates, we multiply the contribution from each round by $r$, which corresponds to dividing by the expected number of rounds in a game.

---

[3] When one decides on a certain set of $k$ basic behaviours and uses the FSA respresentation for the composites strategies, it appears that the situation is equivalent to a game with $k$ actions and with some randomness in the payoffs. This is correct only for the single round game, but when the game is repeated there is usually more than one basic behaviour that are consistent with the action taken. This means that the opponent's action does not always correspond to a unique behaviour but a set of possible behaviours.

The most direct method for evaluating the expected score of a strategy is to truncate the sum over rounds $t$ at some (large) number. When $r$ is close to one, corresponding to a relatively small expected number of rounds, this is a reasonable method. However, when $r$ is small, it is much more efficient to calculate the expected score $s$ as the solution to a linear equation system, in terms of the transition matrix $\tilde{M}$ of the Markov process, the row vector $\boldsymbol{p}$ of expected payoff to the player in each Markov state, and the column vector $\boldsymbol{x}(0)$ giving the distribution over states in the first round:

$$s = r\boldsymbol{p}^{\mathrm{T}}(1-(1-r)\tilde{M})^{-1}\boldsymbol{x}(0). \tag{1}$$

### *Evolutionary dynamics*

Two processes determine the spread and evolution of strategies in the population. First, simple replication serves as a simple model for both cultural transmission (Fudenberg and Levin 1998, and references therein) and genetic inheritance (Maynard Smith 1982, and references therein). We assume a large population with a constant number of players. In each generation, the players come together in pairwise encounters, resulting in a score for each of the players equal to the sum of the payoffs from the rounds in the game. The expected values of the scores are calculated as described above. Next, each player has a number of offspring, proportional to its score. The offspring have the same strategy as the parent. Finally, part of the population is replaced by choosing randomly among the offspring of the players. In order to keep the number of strategies present in the population at a manageable level, a strategy is considered extinct when its fraction falls below a threshold level (taken to be 0.01 in the simulations). The dynamics seems not to be sensitive to the precise value of the threshold level.

Hence, the frequency $x_i$ of strategy $i$ in the population grows with a rate proportional to the advantage relative to the other strategies in the population. The frequency $x'_i$ of strategy $i$ in the next generation is then

$$x'_i = (1-\delta)x_i + \delta\frac{s_i}{\bar{s}}x_i = x_i + \frac{\delta}{\bar{s}}(s_i - \bar{s})x_i, \tag{2}$$

where $s_i$ is the expected score of a player with strategy $i$, $\bar{s}$ is the average expected score in the population ($\bar{s} > 0$), and $\delta$ is the fraction of the population replaced in each generation. From the second equality in (2), we see that the frequency of a strategy increases if and only if its expected score exceeds the average score in the population.

---

[4] A detailed description of this process, and of other implementation issues, is available on request from the authors.

Second, in order to explore the space of strategies, diversity is introduced by making small random modifications (mutations) to the strategy of players chosen randomly from the population. A mutation may add or remove states or transitions between states, change order of transitions, conditions for transitions to occur, or the basic behaviour of states. Addition of a state, illustrated in Figure 4, is done so that the play of the strategy is unchanged. This enables a strategy to grow incrementally. The mutation rate is low, of the order of one mutation per 1000 generations per individual. If the mutation rate is too high, the population may cross into a regime where selection cannot keep pace with the rate at which new variants enter the population. When this happens, selection breaks down.

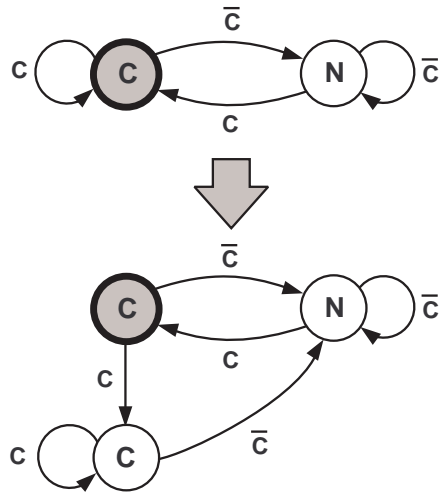

**Figure 4:** There are several different ways to change a strategy by mutations: by adding or removing states, changing the connectivity of the states, or by changing the actions in the states. Shown here is addition of a new state, by duplicating the grey state in such a way that the strategy's behaviour is unchanged. A new node is created, and all the edges leaving the original node are copied to this node. All edges of the original state pointing to itself are redirected to the new state. Finally, a match-all edge to the new state is added last in the rule list of the original state. The states are represented as circles, with a label indicating the basic behaviour followed in that state: C for MaxCoop and N for NashSeek. A thick border indicates the initial state. The transitions between states are represented as arrows pointing from the old state to the new, with the condition for each transition are shown at the corresponding arrow: a C corresponds to requiring a positive match to MaxCoop, and a $\overline{C}$ to requiring a negative match. Note also that the strategy is a simple example of a reciprocating strategy, analogous to the well-known Tit-For-Tat strategy in the PD literature (Axelrod 1984).

In the first generation of each evolution experiment, we take the population to consist of equal fractions of the basic behaviours, with no composite strategies. Through mutations and selection, the strategies in the population increase in complexity. Among the mutations, addition of states and transitions between states occur as frequently as deletions (except when the size of the FSA is at either end of the range), so that the increase in complexity should correspond to a need for more complex strategies in the evolutionary dynamics. The computational demands increase rapidly with the number of nodes in the FSA. We usually limit the strategies to 10 nodes and 10 transitions from each state. This allows for very complex strategies, and increasing these numbers made no qualitative difference on the strategies that evolve.

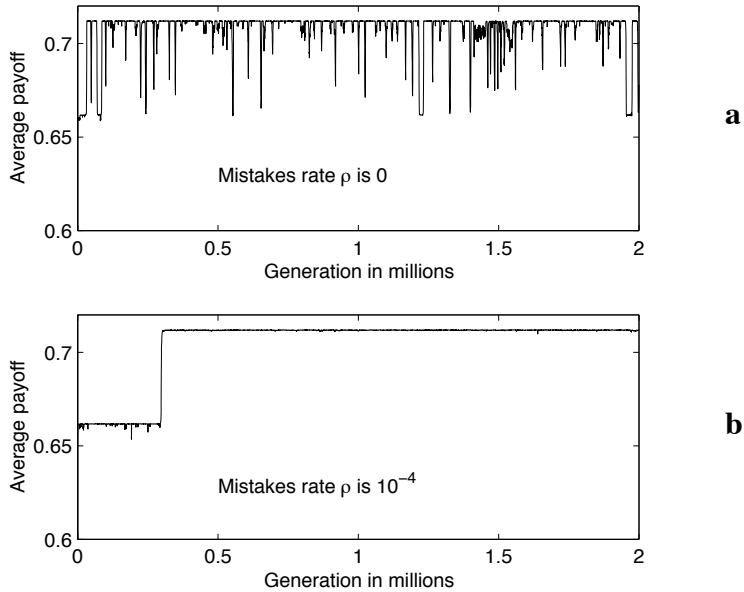

**Figure 5:** Typical runs without mistakes (5a), and with mistakes (5b). With mistakes, it is harder for the mutations to find a solution to the problem of cooperating by playing MaxCoop when possible, while evading exploiting strategies. The mutations of the resulting strategies are much less likely to be fit, and this imply that it is difficult for exploitable strategies to enter the population, as happens when there are no mistakes.

## Results

Qualitatively, our findings are consistent with earlier evolutionary game theory experiments. For instance, we find that cooperation emerges for sufficiently low discount rates, as reciprocating strategies evolve.

Figure 5a shows a typical example of the time evolution of the average payoff in the population. At first, the population consists only of players with basic behaviours. The NashSeek behaviour quickly comes to dominate the population. An analysis of the expected payoffs of the basic behaviours reveals the cause of this. When players have the AntiDiscriminate strategy, the opponents with the Discriminate behaviour receive the highest expected payoff. When a player has the behaviour Sorry, or Generous, the opponent is best off with the behaviour Optimistic, and otherwise the opponent will get the highest expected payoff for the NashSeek strategy. From the above, we see that players with AntiDiscriminate would tend to be replaced by players with the Discriminate behaviour, and then by NashSeek. Similarly, players with Sorry, or Generous, would tend be replaced first by Optimistic and then by NashSeek. Hence, after sufficiently long time, we expect the population to be dominated by the NashSeek behaviour.

At a certain time, the first successful reciprocator appear, which leads to both a rapid increase in the average payoff of the population and to an increase in the complexity of the strategies. This is however unstable; at high levels of cooperation, simple cooperating players, lacking the ability to punish players that deviate from the cooperative behaviour effectively, increase in relative frequency by genetic drift. At a later stage, exploiting strategies may enter the population at the expense of the simple cooperating strategies. The simple cooperative strategies quickly goes extinct and the remaining reciprocating players can then take over again. This leads to a cyclic behaviour, where nevertheless the average amount of cooperation is very high. In Figure 6 we illustrate the cyclic dynamics by showing how the fraction of reciprocating players in the population, defined as players that can both punish NashSeek players and cooperate with MaxCoop players, in Figure 6a correlates with the emergence of exploiting players and the dips in the average score in the population (Figure 6b).

The evolved strategies deal effectively with deviations from cooperation. The simplest such strategies compare the opponents action to that of MaxCoop, and if the actions are not the same the player switches to a more non-cooperative strategy, mostly NashSeek, and stay with this strategy for the rest of the game. The NashSeek behaviour is favoured as punishment behaviour over, for instance, the Punish strategy, because in most populations NashSeek is less costly for the punisher than other alternatives. This composite strategy can be viewed as a trigger strategy, which represents a simple mechanism for establishing cooperation in many situations (Schelling 1978).

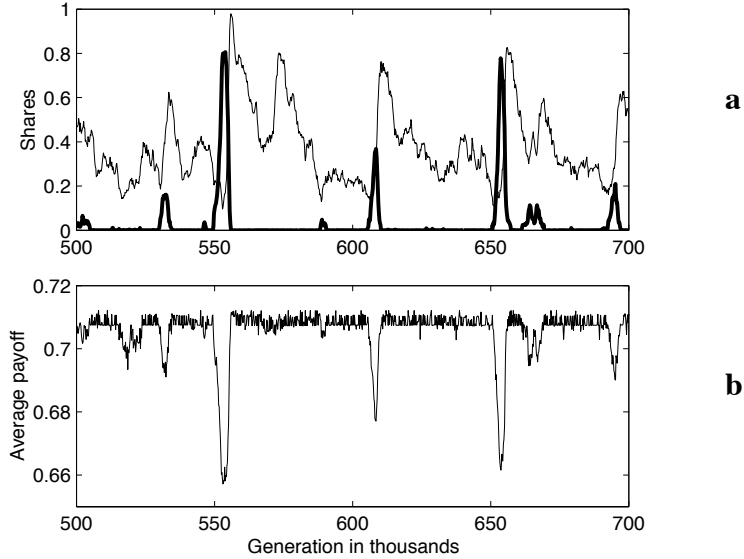

**Figure 6:** The figure illustrates how non-punishing strategies take over from cooperating strategies with punishment, and the consequent collapses due to exploiting strategies, in the run shown in Figure 5a. For clarity, only a short time period is shown here. In the top (6a) is shown the share of players that play like MaxCoop when meeting others with the same behaviour but punish those that deviate from MaxCoop (thin line) and the share of exploiters (thick line). The bottom part (6b) shows the average payoff of the population. When the average payoff of the population is high, the share of strategies with a punishment mechanism decreases, due to an increase of mutant strategies without such a mechanism. Exploiting strategies are held back by the punishments. When the share of punishing strategies fall below a critical level, exploiters get payoffs above the population average and quickly comes to dominate the population. This drastically lowers the population average payoff, which gives the cooperate-and-punish strategies the edge, so that they can take over the population. This cycle then repeats.

The choice of leaving cooperation forever at the first deviation is an effective but severe form of reciprocation. If the players make occasional mistakes, such strategies can be expected to do far worse, in long games, than players with a mechanism for resuming cooperation. On the other hand, resuming cooperation must be done in a way that prevents exploitation. In order to test this scenario, we introduced mistakes into the game model as a probability that, in each round, players take the opposite action to the one intended. We extend the transition rules to include conditions on the player's own actions (in terms of behaviours), in addition to those of the opponent. The resulting game can still be expressed as a Markov process, but with a modified transition matrix.

At large mistake rates, cooperative strategies no longer emerge. In these situations, the population becomes dominated by the more individualistic NashSeek behaviour. Small but positive mistake rates, however, leads to the emergence of more robust strategies. This is illustrated in Figure 5, which shows a comparison between two typical runs, without mistakes and with low but positive mistakes rate. The evolved mechanisms for resuming cooperation gives a decrease in payoff for the individuals, compared to simply cooperating, but the long-term average payoff in the population is nevertheless increased. It appears that the mistake leads to a mechanism that does not allow composite strategies to forget punishment. Typically, we find that the punishment mechanisms that evolve include a complicated sequence of punishments, which involves both players deviating from MaxCoop after a single mistake. Therefore, a strategy that by mutation looses its punishment mechanism, e.g., by resuming cooperation directly after a mistake, will be punished anyway and will therefore bear a higher cost than without the mutation. In this way, the mistakes makes it a selective disadvantage to forget the punishment mechanism, and the instability mechanism in the evolutionary dynamics at the cooperative level seen in Figure 5a is removed. Stabilising mechanisms from mistakes have been investigated in the PD game (Molander 1985; Boyd 1989; Lindgren 1992).

**Discussion**

The random payoff game may serve as a more realistic test model for investigation of cooperative behaviour, compared to the PD, as it reflects the non-static and non-symmetric character of real repeated interactions. One of the points with this game is that cooperation is not simply associated with a certain action, but cooperation needs to be understood as a phenomenon of the behaviour in the repeated game. In the game discussed here, it becomes clear that the MaxCoop strategy is a cooperative strategy: for two players using this strategy, one may say that they work together for a common goal – to maximise the sum of both players' payoffs.

Since the MaxCoop strategy can be exploited (by, for example, NashSeek), a higher-level strategy is needed in an evolutionary or social context, in order to cope with any attempt to exploit. For this, we have introduced the composite strategy as a minimal model of a cognitive process of an agent. Despite the simplicity of the model, we think that the combination of interpretation of the opponent's actions in terms of basic behaviours, and change of the own state of mind as a result of the interpretation, can be used as a general cognitive process description in a large number of agent based models of biological and social behaviour.

We model only deterministic strategies. There are, however, two types of uncertainty included in our class of models. First, we have one model modification in which mistakes occur, which means that a player fails to perform the action dictated by the strategy. Secondly, when a player interprets the action of the opponent in terms of basic behaviours, there is the possibility that two behaviours would give the same action. It is then not possible to deduce the opponent's behaviour from the actions and the payoff matrix alone. For example, in 52% of the payoff matrices, strategies MaxCoop and Punish give the same action and are thus indistinguishable.

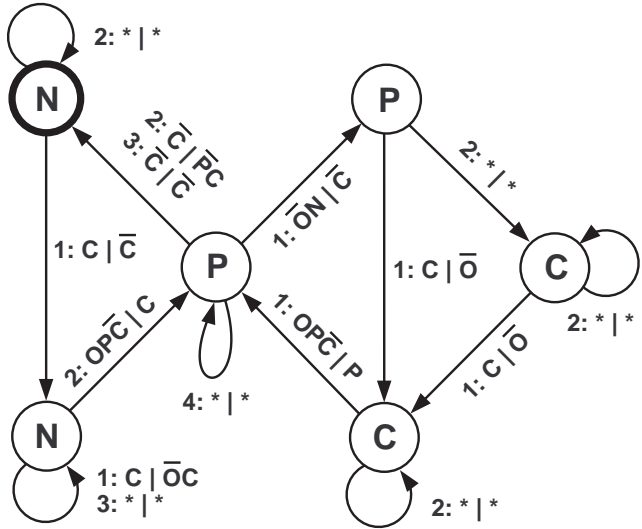

**Figure 7:** This figure shows an evolved strategy that is capable avoiding exploitation, while playing MaxCoop most of the time with cooperating opponents. Two of the internal states (marked with C) correspond to the individual being in a cooperative mood, aiming for the maximum sum of both players' scores. The four other states represent a cognitive process that deals with the situation when the opponent deviates from MaxCoop, in which case the player enters complicated series of actions that punishes the opponent by using a combination of Punishment and NashSeek (marked with P and N, respectively) strategies in a series of following rounds. As in Figure 4, states and transitions between states are shown as circles and arrows, respectively. A thick border indicates the initial state. Each circle has a letter, which gives the basic behaviour followed in that state. In the transition rules, we also find O for Optimistic. The transition rules are more complex, compared to Figure 4, since they include conditions on both players. Each rule is numbered in the order the player considers them. In each round, the first rule to match is followed. Each letter in a rule corresponds to requiring a match between the action of a player and the corresponding basic behaviour for that player. When the letter has a line over it, the action is required not to match the basic behaviour. The letters are presented in two groups, separated by a vertical line, the ones pertaining to the action of the opponent to the left of the line and the ones pertaining to the player's own action to the right. Note that most of the rules only involve a few of the nine strategies, the rest are ignored. An asterisk (*) indicates when the action of one of the players is ignored in a rule.

By conditioning the transitions between states on the action values of several strategies, the strategies increase their ability to adapt to the current payoff matrix; insofar the nine single round strategies discriminate different regions in the space of payoff matrices. This is also, to some extent, a simple model of players that think about the opponent's actions and motives in terms of its own; this kind of projection is something humans often do, even when interacting with non-humans like animals or computers.

The results from the evolutionary simulations show that cooperation may be established in the population as a suitable composite strategy is formed that is both cooperative and punishes exploiting strategies. One can see this as an illustration of that the simple results from the repeated PD and the corresponding strategy in that context, Tit-for-tat (that mimics the opponent's previous action), also holds in this much more general situation.

An illustration of how the composite strategy may look like after the evolution of the combined cooperative and punishing response behaviour is shown in Figure 7. Two of the internal states (marked with C) correspond to the individual being in a cooperative mood, i.e., MaxCoop, aiming for the maximum sum of both players' scores. The four other states represent a cognitive process that deals with the situation when the opponent deviates from MaxCoop, in which case the player enters a complicated series of actions that punishes the opponent by using a combination of Punishment and NashSeek strategies in a series of following rounds.

One characteristic feature of the simulations, in the absence of mistakes, is that cooperation is not stable. If the whole population is cooperating, players do not need the punishment mechanism. Therefore, evolutionary loss of memory by mutation leads to an increasing number of individuals incapable of punishment. When this loss is larger than a critical limit, a mutant strategy exploiting the cooperative behaviour may invade. The exploiters then expand at the expense of the pure cooperators, but after that the remaining cooperators with punishment take over again and the cycle starts again. To put it in the words of George Santayana (1905): "Those who cannot remember the past are condemned to repeat it."

**Acknowledgement**

The authors thank one of the reviewers for comments on the manuscript.